\documentclass[preprint]{aastex}
\slugcomment{OU-TAP 115: ApJ in press}

\shorttitle{Variation of Gas Mass Distribution in Galaxy Clusters}

\begin{document}

\title{The Variation of Gas Mass
Distribution in Galaxy Clusters: Effects of Preheating and Shocks}

\author{Yutaka Fujita\altaffilmark{1} and Fumio Takahara}
\affil{Department of Earth and Space Science, Graduate School of
Science, Osaka University, Machikaneyama-cho, Toyonaka, 
Osaka, 560-0043}

\altaffiltext{1}{JSPS Research Fellow}

\begin{abstract}
 We investigate the origin of the variation of the gas mass fraction in
 the core of galaxy clusters, which was indicated by our work on the
 X-ray fundamental plane. Applying a spherical collapse model of cluster
 formation and considering the effect of shocks on preheated
 intracluster gas, we construct a simple model to predict the spatial
 gas distribution of clusters. As is suggested by our previous work, we
 assume that the core structure of clusters determined at the cluster
 collapse has not been much changed after that. The adopted model
 supposes that the gas distribution characterized by the slope parameter
 is related to the preheated temperature. Comparison with observations
 of relatively hot ($\gtrsim 3$ keV) and low redshift clusters suggests
 that the preheated temperature is about 0.5-2 keV, which is higher than
 expected from the conventional galactic wind model and possibly
 suggests the need for additional heating such as quasars or
 gravitational heating on the largest scales at high redshift. The
 dispersion of the preheated temperature may be attributed to the
 gravitational heating in subclusters. We calculate the central gas
 fraction of a cluster from the gas distribution, assuming that the
 global gas mass fraction is constant within a virial radius at the time
 of the cluster collapse. We find that the central gas density thus
 calculated is in good agreement with the observed one, which suggests
 that the variation of gas mass fraction in cluster cores appears to be
 explained by breaking the self-similarity in clusters due to preheated
 gas. We also find that this model does not change major conclusions on
 the fundamental plane and its cosmological implications obtained in
 previous papers, which strongly suggests that not only for the dark
 halo but also for the intracluster gas the core structure preserves
 information about the cluster formation.
\end{abstract}


\section{Introduction}

Correlations among physical quantities of clusters of galaxies are very
useful tools for studying the formation of clusters and cosmological
parameters. Recently, we have found that clusters at low redshifts
($z\lesssim 0.1$) form a plane (the fundamental plane) in the three
dimensional space represented by their core structures, that is, the
central gas density $\rho_{\rm gas,0}$, core radius $r_c$, and X-ray
temperature $T_{\rm gas}$ \citep[Paper~I]{fuj99a}. On the other hand, a
simple theoretical model of cluster formation predicts that clusters
should be characterized by the virial density $\rho_{\rm vir}$ (or the
collapse redshift $z_{\rm coll}$) and the virial mass $M_{\rm vir}$
(\citealt{fuj99b}, Paper~II). Thus, assuming the similarity of the dark
matter distributions, clusters should form a plane in the three
dimensional space of the dark matter density in the core $\rho_{\rm DM,
c}$, the core radius of dark matter distribution $r_{\rm DM, c}$, and
the virial temperature $T_{\rm vir}$ \footnote{In Paper~I, we used the
terms ' virial density', 'virial radius', and ' virial mass' to denote
the dark matter density in the core, the core radius of dark matter
distribution, and the core mass, respectively. This is because we
assumed that the dark matter density in the core is proportional to the
average dark matter density over the whole cluster (Paper II). To avoid
possible confusions, here we use the term ' dark matter', and the term '
virial' will be used to represent spatially averaged quantities of
gravitational matter (mostly dark matter) within the virialized
region.}. However, the relations between the two planes are not simple;
for example, it is found that $\rho_{\rm gas, 0}$ is not proportional to
$\rho_{\rm DM, c}$. In Paper~I, we found that the ratio $\rho_{\rm
gas,0}/\rho_{\rm DM, c}$ is not constant but obeys the relation of
$\rho_{\rm gas,0}/\rho_{\rm DM, c} \propto \rho_{\rm DM, c}^{-0.1}M_{\rm
DM, c}^{0.4}$, where $M_{\rm DM, c}$ is the core mass.
This raises the question how the segregation between gas and dark
matter occurs.

In the hierarchical structure formation, dark halos are expected to obey
scaling relations. In fact, numerical simulations suggest that the
density distribution in dark halos take a universal form as claimed by
\citet{nav96, nav97}. On a cluster scale, it can be approximated by
$\rho_{\rm DM}(r)\propto r^{-2}$ for $r\lesssim 1$ Mpc, where detailed
X-ray observations have been done \citep{mak98}. On the contrary,
observations show that the slope of the density profile of the hot
diffuse intracluster gas has a range of value. Radial surface brightness
profiles of X-ray emission are often fitted with the conventional
$\beta$ model as
\begin{equation}
\label{eq:sur}
I(R)=\frac{I_0}{(1+R^2/r_c^2)^{3\beta_{\rm obs}-1/2}}\;,
\end{equation}
where $\beta_{\rm obs}$ is the slope parameter \citep{cav78}. If the
intracluster gas is isothermal, equation (\ref{eq:sur}) corresponds to
the gas density profile of
\begin{equation}
\label{eq:gas_obs}
\rho_{\rm gas}(r)=\frac{\rho_{\rm gas, 0}}{(1+r^2/r_c^2)
^{3\beta_{\rm obs}/2}}\:.
\end{equation}
Observations show that the slope parameter takes a range $\beta_{\rm
obs} \sim 0.4-1$ \citep{jon84, jon99}. This means that for $r>>r_{\rm
c}$, the density profiles range from $\propto r^{-1.2}$ to $\propto
r^{-3}$, which are more diverse than those of dark matter. Moreover,
observations show that the clusters with large $r_c$ and $T_{\rm gas}$
tend to have large $\beta_{\rm obs}$ (e.g. \citealt{neu99};
\citealt{hor99}; \citealt{jon99}). Since the average gas fraction of
clusters within radii much larger than $r_{\rm c}$ should be universal
and the dark matter distribution of clusters is also universal, the
variation of $\beta_{\rm obs}$ is expected to correlate with that of the
gas fraction in the core region. In other words, the gas fraction at the
core is not the same as that of the whole cluster and is not
proportional to the dark matter density. This fact must be taken care of
when we discuss cosmological parameters using observational X-ray
data. Since the emissivity of X-ray gas is proportional to $\rho_{\rm
gas}^2$, most of the X-ray emission of a cluster comes form the central
region where $\rho_{\rm gas}$ is large. Although in Papers I and II, we
did not take account of the effects of $\beta_{\rm obs}$, we did find
the gas mass fraction in the core region is diverse by analyzing the
X-ray emission. In this paper, we reanalyze the data taking account of
$\beta_{\rm obs}$ and discuss the relation between core and global gas
mass fractions. We will also show that major conclusions on the
fundamental relations are not changed.

The variation of gas mass fraction itself has been investigated by
several authors (e.g. \citealp{ett99, arn99}). \citet{ett99} argue that
it is partially explained if the dark matter has a significant baryonic
component. Another possible explanation of the diverse gas distributions
is that intracluster gas had already been heated before the collapse
into the cluster; the energetic winds generated by supernovae are one
possible mechanism to increase gas entropy (e.g. \citealp{dek86,
mih94}). In fact, \citet{pon99} find that the entropy of the
intracluster gas near the center of clusters is higher than can be
explained by gravitational collapse alone. In order to estimate the
effect of the preheating on intracluster gas, we must take account of
shocks forming when the gas collapses into the cluster; they supply
additional entropy to the gas. \citet{cav97, cav98, cav99} have
investigated both the effects and predicted the relation between X-ray
luminosities and temperatures ($L_{\rm X}-T_{\rm gas}$ relation). They
predicted that the gas distributions of poor clusters are flatter than
those of rich clusters, which results in a steeper slope of $L_{\rm
X}-T_{\rm gas}$ relation for poor clusters. This is generally consistent
with the observations. It is an interesting issue to investigate whether
this scenario provides a natural explanation for the observed dispersion
of gas mass fraction in the cluster core and whether it reproduces the
X-ray fundamental plane we found in Paper I in our general theoretical
scenario.

In order to clarify what determines the gas distribution, we construct
as a simple model as possible. Although many authors have studied the
preheating of clusters
(\citealt{kai91,evr91,met94,bal99,kay99,wu99,val99}), this is the first
time to consider the influence of the preheating and shocks on the
fundamental plane and two-parameter family nature of clusters paying
attention to the difference between the collapse redshift $z_{\rm coll}$
and the observed redshift $z_{\rm obs}$ of clusters explicitly. In
\S\ref{sec:model}, we explain the model of dark matter potential and
shock heating of intracluster gas. In \S\ref{sec:result}, we use the
model to predict $\beta_{\rm obs}-T_{\rm gas}$ and $\beta_{\rm
obs}-r_{\rm c}$ relations, and the fundamental plane and band of
clusters. The predictions are compared with observations.

\section{Models}
\label{sec:model}
\subsection{Dark Matter Potential}
\label{sec:pot}

In order to predict the relations among parameters describing a dark
matter potential, we use a spherical collapse model \citep{tom69,gun72}.
Although the details are described in Paper~II, we summarize them here
for convenience.

The virial density of a cluster $\rho_{\rm vir}$ at the time of the
cluster collapse ($z_{\rm coll}$) is $\Delta_c$ times the critical
density of a universe at $z=z_{\rm coll}$. It is given by
\begin{equation}
  \label{eq:density}
  \rho_{\rm vir} = \Delta_c \rho_{\rm crit}(z_{\rm coll})
           = \Delta_c \rho_{\rm crit,0}E(z_{\rm coll})^2
           = \Delta_c \rho_{\rm crit,0} \frac
             {\Omega_0 (1+z_{\rm coll})^3}
             {\Omega(z_{\rm coll})} \:,
\end{equation}
where $\Omega(z)$ is the cosmological density parameter, and $E(z)^2 =
\Omega_0 (1+z)^3/\Omega(z)$, where we do not take account of the
cosmological constant. The index 0 refers to the values at $z=0$. Note
that the redshift-dependent Hubble constant can be written as $H(z) =
100 h E(z) \rm\; km\; s^{-1}\; Mpc^{-1}$. We adopt $h=0.5$ for numerical
values. In practice, we use the fitting formula of \citet{bry98} for the
virial density:
\begin{equation}
  \label{eq:delta}
  \Delta_c = 18\pi^2 + 60 x - 32 x^2  \;,
\end{equation}
where $x = \Omega(z_{\rm coll})-1$. 

It is convenient to relate the collapse time in the spherical model with
the density contrast calculated by the linear theory. We define the
critical density contrast $\delta_c$ that is the value, extrapolated to
the present time ($t=t_0$) using linear theory, of the overdensity which
collapses at $t=t_{\rm coll}$ in the exact spherical model. It is given
by
\begin{eqnarray}
  \label{eq:crit}
  \delta_c(t_{\rm coll}) 
       &=& \frac{3}{2}D(t_0)\left[
           1+\left(\frac{t_\Omega}{t_{\rm coll}}\right)^{2/3}\right]
               \;\;\;\; (\Omega_0 < 1) \\
       &=& \frac{3(12\pi)^{2/3}}{20}
           \left(\frac{t_0}{t_{\rm coll}}\right)^{2/3}
               \;\;\;\; (\Omega_0 = 1) 
\end{eqnarray}
\citep{lac93}, where $D(t)$ is the linear growth factor given by
equation (A13) of \citet{lac93} and $t_\Omega = \pi
H_0^{-1}\Omega_0(1-\Omega_0)^{-3/2}$.

For a power-law initial fluctuation spectrum $P(k)\propto k^n$, the rms
amplitude of the linear mass fluctuations in a sphere containing an
average mass $M$ at a given time is $\delta \propto M^{-(n+3)/6}$. Thus,
the virial mass of clusters which collapse at $t_{\rm coll}$ is related
to that at $t_0$ as
\begin{equation}
  \label{eq:mass}
  M_{\rm vir}(t_{\rm coll}) 
     =M_{\rm vir, 0}\left[
  \frac{\delta_c(t_{\rm coll})}{\delta_c(t_0)}\right]^{-6/(n+3)}
                  \;.
\end{equation}
Here, $M_{\rm vir, 0} (=M_{\rm vir}[t_0])$ is regarded as a variable
because actual amplitude of initial fluctuations has a distribution.  We
relate $t=t_{\rm coll}$ to the collapse or formation redshift $z_{\rm
coll}$, which depends on cosmological parameters. Thus, $M_{\rm vir}$ is
a function of $z_{\rm coll}$ as well as $M_{\rm vir, 0}$. This means
that for a given mass scale $M_{\rm vir}$, the amplitude of initial
fluctuations takes a range of value, and spheres containing a mass of
$M_{\rm vir}$ collapse at a range of redshift. In the following, the
slope of the spectrum is fixed at $n = - 1$. It is typical of the
scenario of standard cold dark matter for a cluster mass range, and is
consistent with observations as shown in Paper~II.

The virial radius and temperature of a cluster are then calculated by
\begin{equation}
  \label{eq:rad}
  r_{\rm vir} = \left(\frac{3M_{\rm vir}}
               {4\pi \rho_{\rm vir}}\right)^{1/3} \:,
\end{equation}
\begin{equation}
  \label{eq:temp}
  T_{\rm vir} = \gamma \frac{\mu m_{\rm H}}{3 k_{\rm B}}\frac{G M_{\rm
      vir}}{r_{\rm vir}} \:,
\end{equation}
where $\mu (=0.6)$ is the mean molecular weight, $m_{\rm H}$ is the
hydrogen mass, $k_{\rm B}$ is the Boltzmann constant, $G$ is the
gravitational constant, and $\gamma$ is a fudge factor which typically
ranges between 1 and 1.5. In Paper~II, we adopted the value
$\gamma=1$. Note that the value of $\gamma$ is applied only to dark
matter, but not to gas, because we do {\em not} assume $T_{\rm
gas}=T_{\rm vir}$ here. We emphasize that $M_{\rm vir}$, $\rho_{\rm
vir}$, and $r_{\rm vir}$ are the virial mass, density, and radius at the
time of the cluster collapse, respectively. 

\subsection{Shocks and Hydrostatic Equilibrium}
\label{sec:shock}

To study the effect of preheating, we here adopt a very simple model as
a first step. When a cluster collapses, we expect that a shock wave
forms and the infalling gas is heated. In order to derive the postshock
temperature, we use a shock model of \citet{cav98}. For a given preshock
temperature $T_1$, the postshock temperature $T_2$ can be calculated
from the Rankine-Hugoniot relations. Assuming that the shock is strong
and that the shock front forms in the vicinity of $r_{\rm vir}$, it is
approximately given by
\begin{equation}
\label{eq:shock}
k_{\rm B}T_2 = -\frac{\phi(r_{\rm vir})}{3}
+\frac{3}{2}k_{\rm B}T_1\:
\end{equation}
\citep{cav98}, where $\phi(r)$ is the potential at $r$. According to the
virial theorem and the continuity when $T_1$ approaches zero, we should
take $-\phi(r_{\rm vir})/3=k_{\rm B}T_{\rm vir}$. For $r<r_{\rm vir}$,
we assume that the gas is isothermal and hydrostatic, and that the
matter accretion after the cluster collapse does not much change the
structure of the central region of the cluster significantly,
as confirmed by numerical simulations (e.g. \citealt{tak98}). It is to
be noted that even if the density profile of dark matter is represented
by the universal profile \citep{nav96, nav97}, it is not inconsistent
with the isothermal $\beta$ model of gas represented by equation
(\ref{eq:gas_obs}) (\citealt*{mak98, eke98}) within the present
observational scopes. On these assumptions, the gas temperature in the
inner region of a cluster is $T_{\rm gas}=T_2$, and the mass within $r$
of the cluster center $M_{\rm DM}$ is related to the density profile of
intracluster gas, $\rho_{\rm gas}$, by
\begin{equation}
\label{eq:MDM}
M_{\rm DM}(r)= -\frac{k_{\rm B}T_{\rm gas}}{\mu m_{\rm H}G}\; r
\; \frac{d \ln \rho_{\rm gas}}{d \ln r} \;.
\end{equation}
Since $M_{\rm DM}(r_{\rm vir})=M_{\rm vir}$, equations (\ref{eq:temp})
and (\ref{eq:MDM}) yield
\begin{equation}
\label{eq:temp2}
T_{\rm vir} = -\frac{\gamma}{3}\: T_{\rm gas}
\left.\frac{d \ln \rho_{\rm gas}}{d \ln r}\right|_{r=r_{\rm vir}} \;.
\end{equation}
Defining $\beta=T_{\rm vir}/T_{\rm gas}$, the gas density profile is
thus given by
\begin{equation}
\label{eq:beta}
\rho_{\rm gas}(r)\propto r^{-3\beta/\gamma}\:,
\end{equation}
as long as ($d \ln \rho_{\rm gas}/d \ln r$) is nearly constant.

Equation (\ref{eq:shock}) shows that in this model $\beta$ is a function
of only $T_{\rm vir}$ when $T_1$ is regarded as an external parameter,
that is,
\begin{equation}
\label{eq:beta2}
\beta=\frac{T_{\rm vir}}{T_{\rm vir}+(3/2)T_1}\:.
\end{equation}
Since $T_{\rm vir}=T_{\rm gas}\beta$, equation (\ref{eq:beta2}) is
written as
\begin{equation}
\label{eq:beta3}
\beta=\frac{T_{\rm gas}-(3/2)T_1}{T_{\rm gas}}\:.
\end{equation}
Thus, the $\beta-T_{\rm gas}$ relation can be used to determine $T_1$ by
comparing with the observation. Since both $T_{\rm vir}$ and $r_{\rm
vir}$ are the two-parameter families of $z_{\rm coll}$ and $M_{\rm vir,
0}$ (equations [\ref{eq:density}], [\ref{eq:mass}],[\ref{eq:rad}] and
[\ref{eq:temp}]), equation (\ref{eq:beta2}) shows that $\beta$ can be
represented by $r_{\rm vir}$ as $\beta=\beta(r_{\rm vir}, M_{\rm vir,
0})$, if $T_1$ is specified. Recent numerical simulations suggest that
the structure of central region of clusters is related to $z_{\rm coll}$
\citep{nav97}, and in particular $r_{\rm DM, c}$ is proportional to
$r_{\rm vir}$ \citep{sal98,mak98}. Therefore, if we assume that $r_{\rm
DM, c}=r_{\rm c}$ and that $r_{\rm vir}/r_{\rm c}$ is constant as in
Paper~II, $T_1$ can also be determined by comparing the theoretical
prediction of the $\beta-r_{\rm c}$ relation with the observation. Since
a spherical collapse model predicts $r_{\rm vir}(z_{\rm coll}=0) \sim 4$
Mpc and observations show that $r_{\rm c}(z_{\rm coll}=0)\sim 0.5$ Mpc
(Figure 1b in Paper~II), we adopt $r_{\rm vir}/r_{\rm c}=8$ from now on. 
Thus, we obtain $\beta=\beta(8 r_{\rm c}[z_{\rm coll}, M_{\rm vir, 0}],
M_{\rm vir, 0})$.

\section{Results and Discussion}
\label{sec:result}
\subsection{$\beta-T_{\rm gas}$ and $\beta-r_{\rm c}$ Relations}

Using the model constructed in \S\ref{sec:shock}, we predict relations
between $\beta$ and $T_{\rm gas}$, and between $\beta$ and $r_{\rm
c}$. 

\begin{figure}
\figurenum{1}
\epsscale{0.80}
\plotone{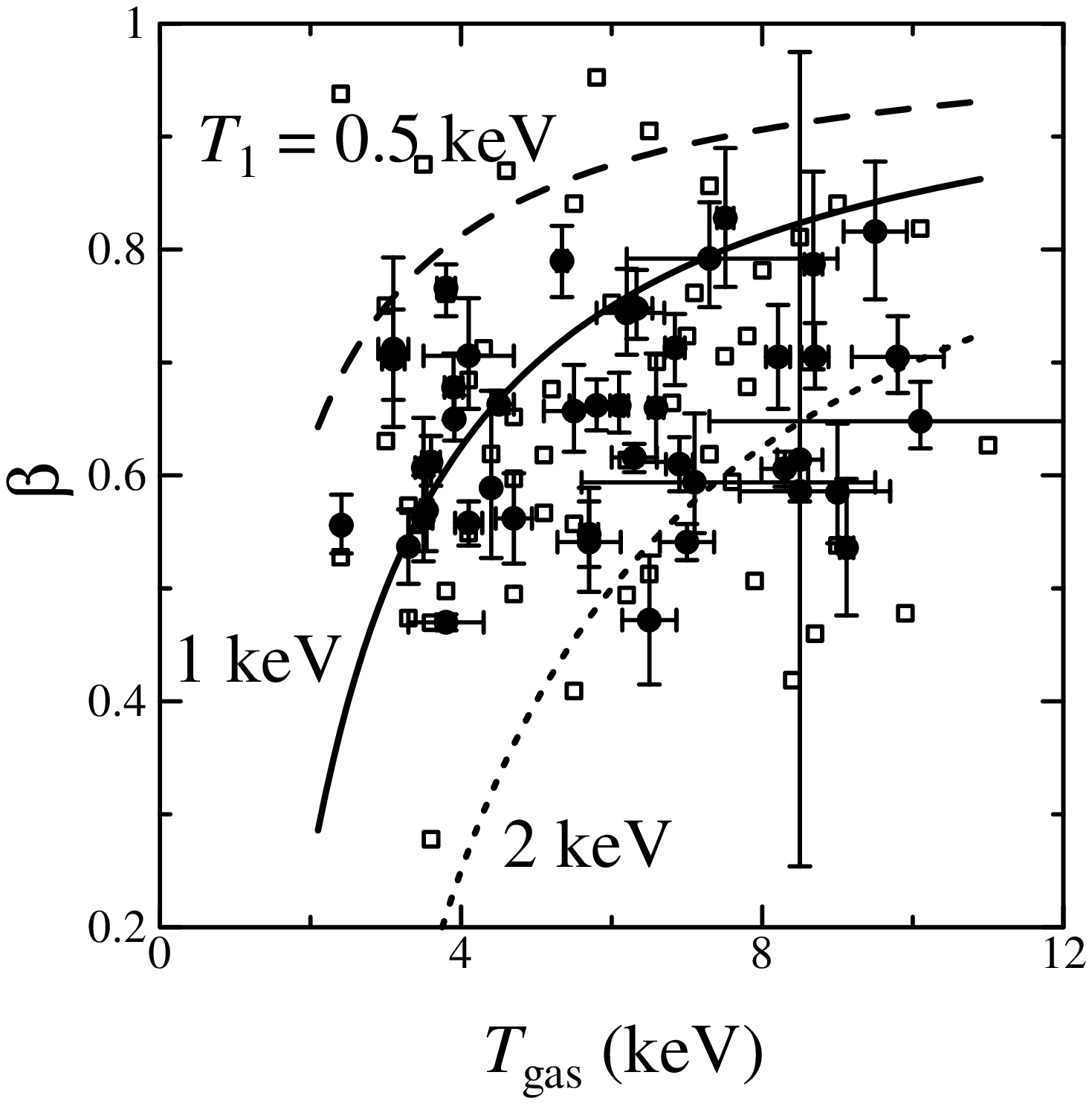}
\caption{The relation between $\beta$ and $T_{\rm gas}$ when
$\gamma=1$. Dashed line: $T_1=0.5$ keV. Solid line: $T_1=1$ keV. Dotted
line: $T_1=2$ keV. The observational data obtained by \citet{moh99}
(filled circles) and \citet{per98} (open squares) are overlaid.
\label{fig1}}
\end{figure}

If $T_1$ is mainly determined by the energetic winds generated in the
forming galaxies or quasars before the formation of clusters, $T_1$
should be constant if subsequent adiabatic heating or cooling is
neglected. However, if, besides the winds, the gravitational energy
released in the subclusters, which later merged into the observed
clusters, contributes to $T_1$, we expect that $T_1$ has a distribution
produced by different merging histories. In order to determine the
distribution in detail, we must calculate the merging histories by Monte
Carlo realizations as \citet[1998]{cav97} did. In this study, however,
we consider the scatter by investigating a range of $T_1$ for
simplicity. We show in Figure \ref{fig1} the $\beta-T_{\rm gas}$
relation for $T_1=0.5$, 1, and 2 keV. The observational data are
overlaid. Since equation (\ref{eq:gas_obs}) is approximated to be
$\rho_{\rm gas}(r)\propto r^{-3\beta_{\rm obs}}$ for $r>>r_c$, the
relation
\begin{equation}
\label{eq:beta_beta}
\beta = \gamma \beta_{\rm obs}
\end{equation}
is obtained by comparing the relation (\ref{eq:beta}).  Thus, in the
following figures, the observed values of $\beta_{\rm obs}$ are
converted by relation (\ref{eq:beta_beta}). In Figure~\ref{fig1} we
assumed $\gamma=1$. As the data, we use only relatively hot ($T_{\rm
gas}\gtrsim 3$ keV) and low redshift ($z\lesssim 0.1$) clusters obtained
by \citet{moh99} and \citet{per98}. Instead of $\beta_{\rm obs}$,
\citet{per98} present velocity dispersions corresponding to
gravitational potential well, $\sigma_{\rm deproj}$, derived with the
deprojection method, ignoring velocity dispersion anisotropies and
gradients . Thus, for the data we assume that $k_{\rm B}T_{\rm vir}=\mu
m_{\rm H}\sigma_{\rm deproj}^2$ and define $\beta$ as $T_{\rm
vir}/T_{\rm gas}$. Figure~\ref{fig1} shows that the observational data
are consistent with $0.5\lesssim T_1 \lesssim 2$ keV but it seems that a
single value of $T_1$ does not represent the range of data. The
preheating ($T_1>0$) is expected to reduce $\beta$ of the clusters with
small $T_{\rm gas}$ (Figure~\ref{fig1}). At first glance, no
correlations between $\beta$ and $T_{\rm gas}$ are recognized
observationally in this temperature range. However, some reports on the
existence of a weak correlation have been made when clusters with lower
$T_{\rm gas}$ are included (e.g. \citealt{hor99}). Thus, our prediction
is not inconsistent with the observations.

\begin{figure}
\figurenum{2}
\epsscale{1.20}
\plottwo{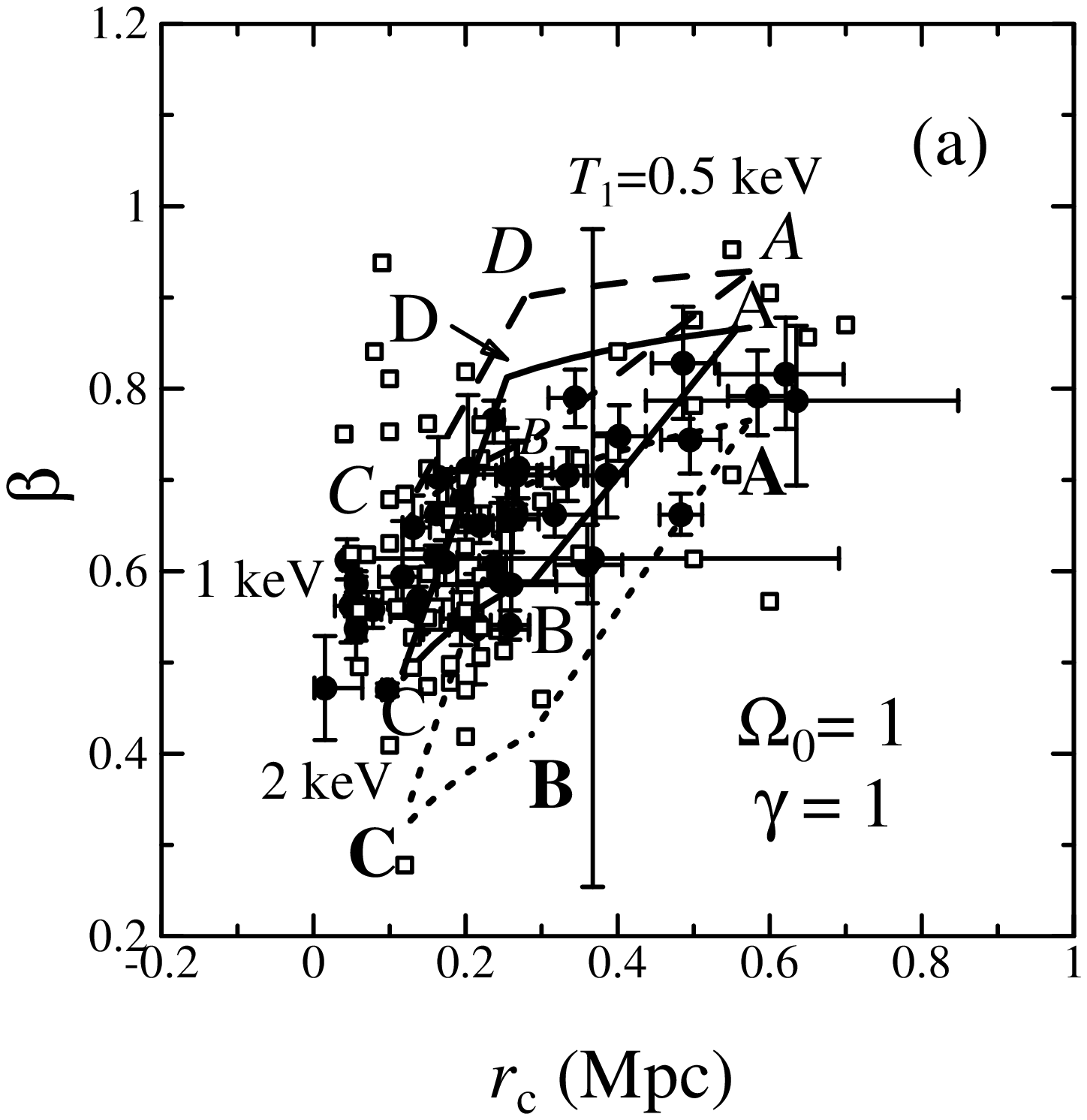}{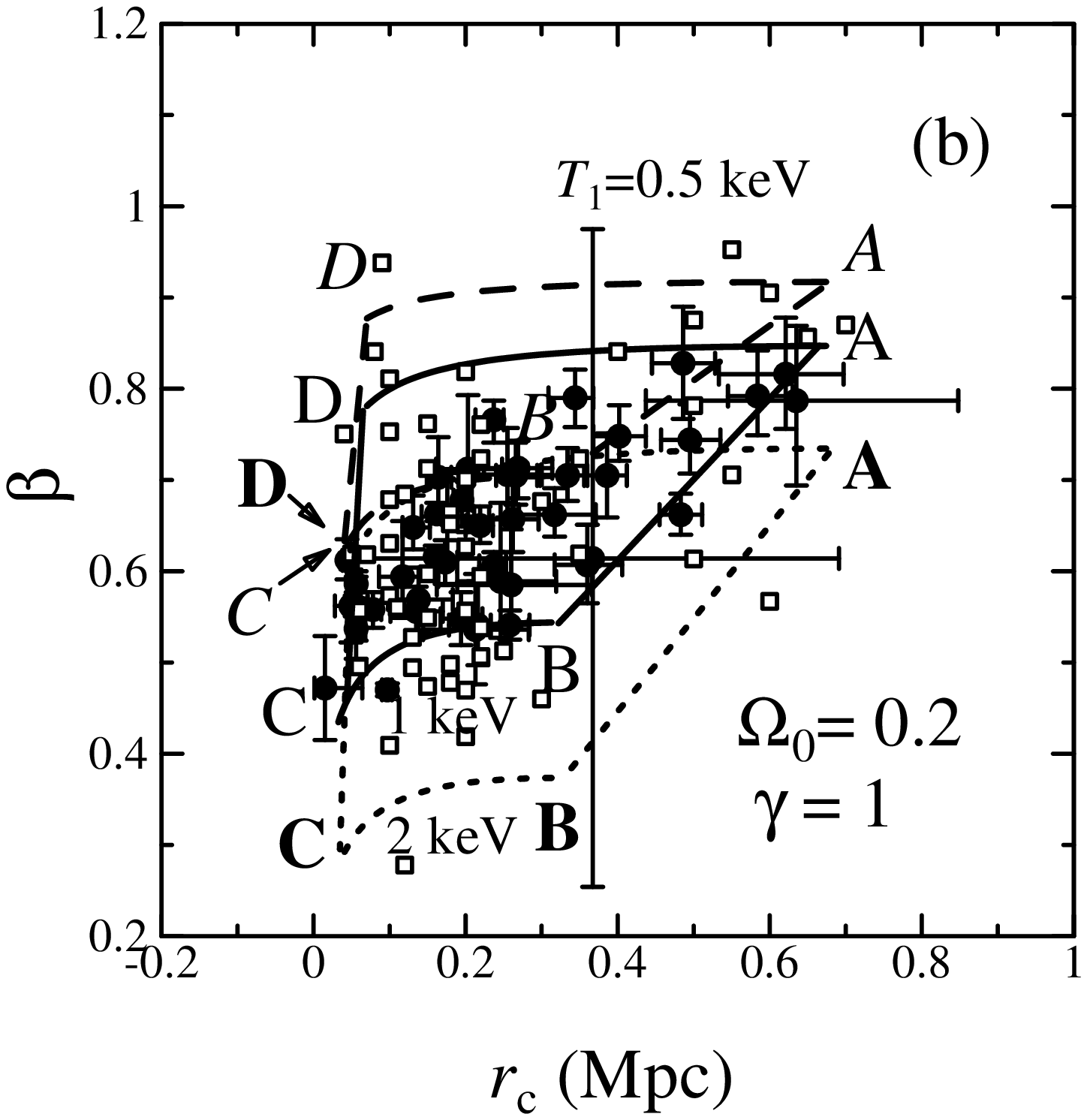}
\caption{Theoretical predictions of $\beta-r_{\rm c}$
relation in the case of (a) $\Omega_0=1.0$ and $\gamma=1$ (b)
$\Omega_0=0.2$ and $\gamma=1$. Dashed line: $T_1=0.5$ keV. Solid line:
$T_1=1$ keV. Dotted line: $T_1=2$ keV. The arcs AD and BC correspond to
$M_{\rm vir, 0}=5\times 10^{15}\rm\; M_{\sun}$ and $5\times 10^{14}\rm\;
M_{\sun}$, respectively. The segments AB corresponds to $z_{\rm coll}=0$
and the segments CD corresponds to $z_{\rm coll}=0.5$ in (a) and $z_{\rm
coll}=4$ in (b). The observational data obtained by \citet{moh99} (filled
circles) and \citet{per98} (open squares) are overlaid.
\label{fig2}}
\end{figure}

As discussed in \S\ref{sec:shock}, the $\beta-r_{\rm c}$ relation is
represented by two parameters $z_{\rm coll}$ and $M_{\rm vir, 0}$, for a
given value of $T_1$. The results are shown in Figure~\ref{fig2} for
$\gamma=1$. Figure~\ref{fig2}a and ~\ref{fig2}b are for $\Omega_0=1$ and
0.2, respectively. For comparison, we also present observational data
\citep{moh99,per98}. As was in Paper~I, for the data of \citet{moh99} we
use here only the component of surface brightness reflecting the global
structure of clusters, although the central component (so-called cooling
flow component) may also have formed in the scenario of hierarchical
clustering \citep{fuj99c}.

The mass $M_{\rm vir, 0}$ corresponds to the mass of clusters collapsed
at $z\sim 0$ and takes a range of value due to the dispersion of initial
density fluctuation of the universe. Since observations and numerical
simulations show $M_{\rm vir, 0}\sim 10^{15}\;\rm M_{\sun}$
\citep{evr96}, the observational data are expected to lie between the
two lines of $M_{\rm vir, 0}=5\times 10^{14}\;\rm M_{\sun}$ (arc BC) and
$M_{\rm vir, 0}=5\times 10^{15}\;\rm M_{\sun}$ (arc AD) for fixed $T_1$
in Figure~\ref{fig2}. Note that the distribution of $M_{\rm vir, 0}$
degenerates on the lines in Figure~\ref{fig1}. In Figure~\ref{fig2}, the
positions along the arcs AD and BC indicate the formation redshifts of
the clusters. When $\Omega_0 = 1$, most of the observed clusters should
have collapsed at $z\sim 0$ because clusters continue growing even at
$z=0$ \citep{pee80}. Thus, the cluster data are expected to be
distributed along the part of the lines close to the point of $z_{\rm
coll}=0$ (segment AB). In fact, calculations done by \citet{lac93}, and
\citet{kit96} show that if $\Omega_0 = 1$, most of present day clusters
($M_{\rm vir}\sim 10^{14-15}\rm\: M_{\sun}$) should have formed in the
range of $z_{\rm coll}\lesssim 0.5$ (parallelogram ABCD in
Figure~\ref{fig2}a). In contrast, when $\Omega_0 = 0.2$, the growth rate
of clusters decreases and cluster formation gradually ceases at $z
\lesssim 1/\Omega_0-1$ \citep{pee80}. Thus, in Figure~\ref{fig2}b,
cluster data are expected to be distributed between the points of
$z_{\rm coll}=0$ (segment AB) and $z_{\rm coll} = 1/\Omega_0-1$ (segment
CD) and should have a two-dimensional distribution (parallelogram ABCD). 
Thus, compared with the observations, the models in Figure~\ref{fig2}
show that $T_1\sim 1$ keV and $\Omega_0<1$ are preferred. The latter
result is quite consistent with that of Paper II, where we found that
the $T_{\rm gas}-r_{\rm c}$ relation suggests $\Omega_0<1$. Since
$\beta$ is related to $T_{\rm gas}$ by equation (\ref{eq:beta3}), the
$\beta-r_{\rm c}$ relation is equivalent to the $T_{\rm gas}-r_{\rm c}$
relation for a fixed value of $T_1$. Note that in Figure~\ref{fig2}
predicted regions corresponding to different $T_1$ overlap each other;
this implies that the position of a cluster in Figure~\ref{fig2} does
not uniquely correspond to $T_1$ in contrast to Figure~\ref{fig1}. For a
given $\beta$ and $r_{\rm c}$, larger $T_1$ corresponds to larger
$M_{\rm vir, 0}$ or larger amplitude of the initial fluctuation.

The dispersion of $T_1$ appears to be caused by gravitational heating in
subclusters that are to merge to the cluster ($T_{\rm gas}\gtrsim 3$
keV) at the time of the cluster formation. In fact, Figure~\ref{fig2}b
shows that observed clusters are situated close to the line of $M_{\rm
vir, 0}=5\times 10^{15}\;\rm M_{\sun}$ (arc AD) when $T_1\sim 2$ keV,
while they are situated close to the line of $M_{\rm vir, 0}=5\times
10^{14}\;\rm M_{\sun}$ (arc BC) when $0.5<T_1<1$ keV. Moreover,
Figure~\ref{fig1} suggests that clusters with large $T_{\rm gas}$ favor
large $T_1$. These may reflect that clusters with larger (smaller)
$M_{\rm vir, 0}$ or $T_{\rm gas}$ tend to have more (less) massive
progenitors with larger (smaller) $T_1$, although these are only loose
tendencies, and we need more samples and more improved models to obtain
a definite conclusion.

Note that gravitational heating in subclusters itself is a self-similar
process and does not modify self-similar scaling relations such as the
luminosity-temperature relation (e.g. \citealp{eke98}). Thus, an
additional entropy other than expected from purely gravitational
assembly of a cluster must be injected into the gas. \citet{val99}
investigate the entropy evolution of intergalactic medium (IGM) and find
that clusters with $T_{\rm vir}\sim 0.5$ keV are affected by the
additional entropy when it is generated by quasar heating. This is
because the additional entropy is comparable to the entropy generated by
gravitational collapse of the clusters. In other words, the adiabatic
compression of the gas from the preheated IGM alone can heat the gas up
to $T_{\rm ad, cl}\sim 0.5$ keV. Therefore, in addition to the
gravitational processes in subclusters, the preheating may significantly
contribute to $T_1$, and the lower bound of which is given by $T_{\rm
ad, cl}$. If $T_{\rm ad, cl}\sim 0.5$ keV, this is consistent with our
result (Figures \ref{fig1} and \ref{fig2}). \citet{val99} also
investigate the case when only supernova heating is taken into account
and quasar heating is ignored. The result is $T_{\rm ad, cl}< 0.1$
keV. In this case, and effects of the preheating is small and we expect
that $\beta$ does not much depend on $T_{\rm gas}$ and $r_{\rm c}$,
although $\beta$ would have a scatter owing to the difference of merging
history. This is inconsistent with the observations. The insufficient
power of the supernova heating is also suggested by \citet{wu99}
\citep[but see][]{loe99}. Another possible source of heating is that due
to shocks forming at higher redshift on the largest scales, such as
filaments and sheets. \citet{cen99} indicate that most of baryons at low
redshift should have a temperature in the range of $10^5-10^7$ K. The
relatively large value of $T_1$ may reflect this temperature.

\begin{figure}
\figurenum{3}
\epsscale{0.55}
\plotone{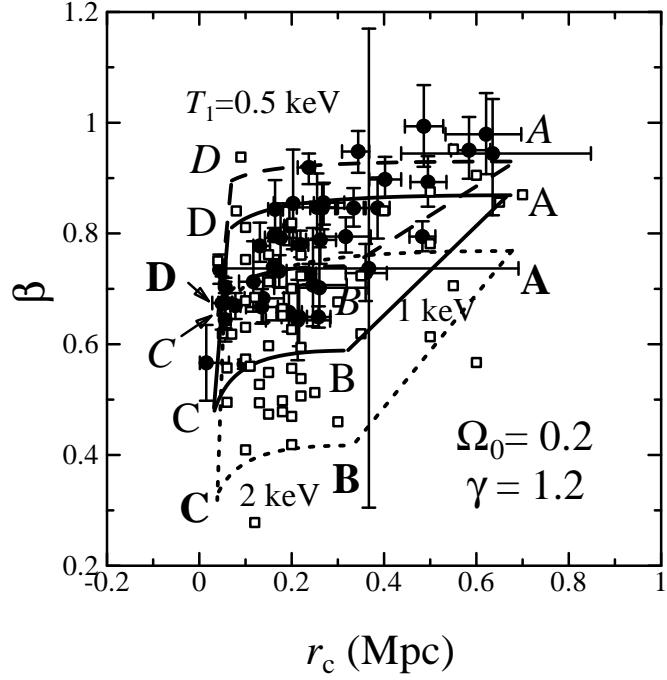}
\caption{The same as Figure~\ref{fig2}b but for
$\gamma=1.2$ \label{fig3}}
\end{figure}

We also investigate the case of $\gamma=1.2$ and $\Omega_0=0.2$, which
are presented in Figure~\ref{fig3}. In this case, the model of $T_1=0.5$
keV is preferred especially for the data obtained by \citet{moh99}. This
means that $\gamma$ and $T_1$ are correlated and they cannot be
determined independently. However, the model of $\gamma> 1.2$ is
inappropriate because $\beta=\gamma\beta_{\rm obs}$ exceeds unity for
some observational data while relation (\ref{eq:beta2}) or
(\ref{eq:beta3}) limits $\beta$ to less than one.
If a cluster is not isothermal, the temperature in the central region
$T_{\rm gas}$ should be larger than $T_2$ \citep{cav98}. In this case,
the discrepancy between the model and the observations is more
significant. Thus, it seems to be difficult to construct a model that
predicts $\beta>1$.

\subsection{The Fundamental Band and Plane}

It is interesting to investigate whether the gas distribution in
clusters derived above is consistent with the observations of central
gas fraction, and the fundamental band and plane we found in
Paper~I. The shapes of the band and plane are also related to the origin
of the observed relation of $L_{\rm X}\propto T_{\rm gas}^3$
(Paper~I). We did not explore the origin of the variation of the central
gas mass fraction in previous papers, where $\beta_{\rm obs}$ was
regarded as constant. Below, we will show that this is related to the
variation of $\beta$.

From relation (\ref{eq:beta}), the gas density at the cluster core is
approximately given by
\begin{equation}
\label{eq:rho_0_ori}
\rho_{\rm gas, 0}
=\rho_{\rm gas}(r_{\rm vir})
(r_{\rm vir}/r_c)
^{3\beta/\gamma}\:,
\end{equation}
where $r_{\rm vir}$ and $\beta(T_{\rm vir}, T_1)$ are functions of
$z_{\rm coll}$ and $M_{\rm vir, 0}$ (\S\ref{sec:model}), and $\rho_{\rm
gas}(r)$ is the gas density at radius $r$ from the cluster center. We
assume that the profile of dark matter is isothermal ($\rho_{\rm
DM}\propto r^{-2}$) at least for $r_{\rm c}\lesssim r \lesssim r_{\rm
vir}$, and $\rho_{\rm DM, c}=64 \rho_{\rm vir}$. Moreover, we assume
that the average gas fraction within radius $r_{\rm vir}$ is $f_{\rm
gas}(r_{\rm vir})=0.25$ regardless of $z_{\rm coll}$ and $M_{\rm vir,
0}$. The value of $f_{\rm gas}$ is nearly the largest gas mass fraction
of observed clusters (e.g. \citealt{dav95, ett99}). On these
assumptions, the central gas density and the gas fraction at the cluster
core are respectively given by
\begin{eqnarray}
\rho_{\rm gas, 0}
 &=& \left(1-\frac{\beta}{\gamma}\right)
 f_{\rm gas}\;\rho_{\rm vir}(z_{\rm coll})
\left(\frac{r_{\rm vir}}{r_{\rm c}}\right)
^{3\beta/\gamma} \nonumber \\
 &=&0.25\left(1-\frac{\beta}{\gamma}\right)
 \rho_{\rm vir}(z_{\rm coll})
\;8^{3\beta/\gamma} 
\:\label{eq:rho_0}
\end{eqnarray}
and
\begin{equation}
f_{\rm gas}(0) = 0.25\left(1-\frac{\beta}{\gamma}\right)
\;8^{3(\beta/\gamma)-2}\label{eq:frac}\:, 
\end{equation}
where $f_{\rm gas}(0)\equiv \rho_{\rm gas, 0}/\rho_{\rm DM, c}$ is the
gas fraction at the cluster center. The above equations are valid when
$\beta<\gamma$. Note that in Paper~II, we derive the central gas density
according to the relation $\rho_{\rm gas, 0}\propto \rho_{\rm vir}f_{\rm
gas}(0)$, in which $f_{\rm gas}(0)$ is differently determined by the
observations\footnote{In Papers~I and II, we assumed that $T_{\rm
gas}=T_{\rm vir}$ and did not take account of the variation of
$\beta_{\rm obs}$ when we derive $f_{\rm gas}(0)$ from observations of
$\rho_{\rm gas, 0}$, $r_{\rm c}$, and $T_{\rm gas}$.}. In contrast, in
equation (\ref{eq:rho_0}), we derive $\rho_{\rm gas, 0}$ assuming that
$f_{\rm gas}(r_{\rm vir})=constant$.

\begin{figure}
\figurenum{4}
\epsscale{0.55}
\plotone{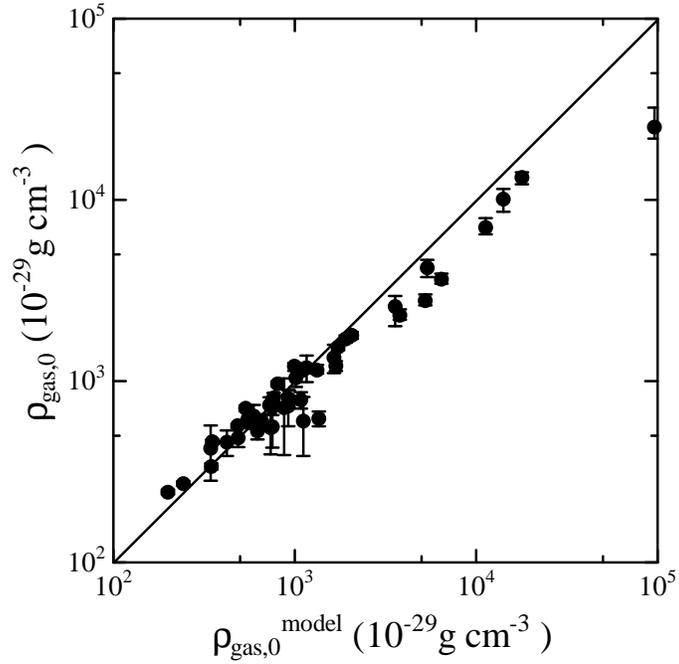}
\caption{The relation between $\rho_{\rm gas, 0}^{\rm model}$
of equation (\ref{eq:rho_model}) and $\rho_{\rm gas, 0}$. The line
corresponds to $\rho_{\rm gas, 0}=\rho_{\rm gas, 0}^{\rm model}$
\label{rhorho}}
\end{figure}

The above model values (equation [\ref{eq:rho_0}] or [\ref{eq:frac}])
can be obtained from observational data. Using equations (\ref{eq:rad}),
(\ref{eq:temp}), and (\ref{eq:beta_beta}) we obtain
\begin{equation}
\label{eq:rho_vir}
\rho_{\rm vir}=\frac{9k_{\rm B}T_{\rm gas}}{4\pi G\mu m_{\rm H}}
\frac{\beta_{\rm obs}}{(8r_{\rm c})^2}
\;,
\end{equation}
where we used the relations of $T_{\rm vir}=\beta T_{\rm gas}$ and
$r_{\rm c}=r_{\rm vir}/8$. Thus, using equation (\ref{eq:beta_beta}),
the right hand of equation (\ref{eq:rho_0}) should be written as
\begin{equation}
\label{eq:rho_model}
\rho_{\rm gas, 0}^{\rm model}\equiv
0.25\beta_{\rm obs}(1-\beta_{\rm obs})
\frac{9k_{\rm B}T_{\rm gas}}{4\pi G\mu m_{\rm H}}
\frac{8^{3\beta_{\rm obs}}}{(8r_{\rm c})^2}
\:.
\end{equation}
Hence, $\rho_{\rm gas, 0}^{\rm model}$ can be derived from the
observable quantities $r_{\rm c}$, $T_{\rm gas}$, and $\beta_{\rm obs}$.
Figure~\ref{rhorho} displays a plot of $\rho_{\rm gas, 0}$ and
$\rho_{\rm gas, 0}^{\rm model}$ based on the data obtained by
\citet{moh99}. Note that \citet{per98} do not present $\rho_{\rm gas,
0}$.  We do not show the uncertainties of $\rho_{\rm gas, 0}^{\rm
model}$ to avoid complexity. Here we use only $\rho_{\rm gas, 0}$
corresponding to the global cluster component as we did in Paper~I.

Figure~\ref{rhorho} shows that $\rho_{\rm gas, 0}$ well agrees with
$\rho_{\rm gas, 0}^{\rm model}$ although $\rho_{\rm gas, 0}$ is slightly
smaller than $\rho_{\rm gas, 0}^{\rm model}$ for clusters with large
$\rho_{\rm gas, 0}^{\rm model}$. Thus, we conclude that the variation of
$f_{\rm gas}(0)$ is due to that of the slope parameter of the gas
distribution $\beta$ within $r_{\rm vir}$. One possible reason for the
slight disagreement between $\rho_{\rm gas, 0}^{\rm model}$ and
$\rho_{\rm gas, 0}$ is an uncertainty of the value of $f_{\rm
gas}(r_{\rm vir})$. Another is the influence of central excess emission
of clusters. When the distance to a cluster is relatively large, the
center and global surface brightness components may not be distinguished
even if the two components exist. In this case, the cluster may be
considered that it has only a global component. However, when the
central emission is strong, the fitting of the surface brightness
profile by one component may be affected by the central emission and may
give a smaller core radius than the real. This may make $\rho_{\rm gas,
0}^{\rm model}$ large for the clusters. In fact, clusters with
$\rho_{\rm gas, 0}^{\rm model}>3\times 10^{-26}\rm\; g\; cm^{-3}$ are
regarded by \citet{moh99} as having only one (global) component of
surface brightness. Note that core radii derived by \citet{per98} may be
less affected by the central emission because they take account of
cooling flows and the gravitation of central cluster galaxies, which are
responsible for the central emission, for {\em all} clusters they
investigate (Figures \ref{fig2} and \ref{FB}b).


We present the theoretically predicted relations among $\rho_{\rm gas,
0}$, $r_{\rm vir}$, and $T_{\rm gas}$ in Figure~\ref{FB}. Although these
relations are presented in Paper~II using the observed relation between
$f_{\rm gas}(0)$ and $M_{\rm DM, c}$, here we plot them by directly
using $\beta$. For lines in Figure~\ref{FB}, we use the relation $T_{\rm
gas}=T_{\rm vir}/\beta$. For comparison, we plot the observational data
in the catalogue of \citet{moh99} and \citet{per98}.  For the data, we
use the relation $r_{\rm vir}=8 r_{\rm c}$. Figure~\ref{FB} shows that
our model can well reproduce the band distribution of observational data
in the ($\rho_{\rm gas, 0}, r_{\rm c}, T_{\rm gas}$)-space. Moreover,
our model can explain the planar distribution of the observational data. 
In Paper~I, we find that the observational data satisfy the relation of
the fundamental plane, $\rho_{\rm gas, 0}^{0.47}r_{\rm c}^{0.65}T_{\rm
gas}^{-0.60}\propto constant$. For $M_{\rm vir, 0}\sim 10^{15}\;\rm
M_{\sun}$ and $z_{\rm coll}\lesssim 2$, our model with $\Omega_0=0.2$
and $T_1=1$ keV predicts the plane of $\rho_{\rm gas, 0}^{0.32}r_{\rm
c}^{0.64}T_{\rm gas}^{-0.70}\propto constant$, which is approximately
consistent with the observation. Note that the index of $\rho_{\rm gas,
0}$ is somewhat smaller than the observed value considering the
uncertainty ($\sim 0.1$), which may be related to the slight
disagreement between $\rho_{\rm gas, 0}^{\rm model}$ and $\rho_{\rm gas,
0}$ (Figure~\ref{rhorho}). The plane is represented by the two
parameters, $z_{\rm coll}$ and $M_{\rm vir, 0}$, as discussed in
\S\ref{sec:model}. Since the cross section of the fundamental plane
corresponds to the observed $L_{\rm X}-T_{\rm gas}$ relation, and the
fundamental plane corresponds to the observed dependence of $f_{\rm
gas}(0)$ on $\rho_{\rm DM, c}$ and $M_{\rm DM, c}$ (Paper~I), our model
can also reproduce these relations. These results strengthen our
interpretation that the difference of gas distribution among clusters is
caused by heating of the gas before the cluster collapse and by shock
heating at the time of the cluster collapse (equation [\ref{eq:beta2}]
or [\ref{eq:beta3}]).

\section{Conclusions}

We have investigated the influence of heating before cluster collapse
and shocks during cluster formation on the gas distribution in the
central region of clusters of galaxies. We assumed that the core
structure has not much changed since the formation of a cluster. Using a
spherical collapse model of a dark halo and a simple shock model, we
predict the relations among the slope of gas distribution $\beta$, the
gas temperature $T_{\rm gas}$, and the core radius $r_{\rm c}$ of
clusters. By comparing them with observations of relatively hot
($\gtrsim 3$ keV) and low redshift clusters, we find that the
temperature of the preheated gas collapsed into the clusters is about
$0.5-2$ keV. Since the temperature is higher than that predicted by a
preheating model of supernovae, it may reflect the heating by quasars or
gravitational heating on the largest scales at high redshift. Moreover,
gravitational heating in subclusters assembled when the clusters formed
also seems to affect the temperature of the preheated gas and produce
the dispersion in the preheating temperature. Assuming that the global
gas mass fraction of clusters are constant, we predict that the gas mass
fraction in the core region of clusters should vary correlating with
$\beta$ through a simple law, which is shown to be consistent with the
observations. Thus, we conclude that the variation of the gas mass
fraction in the cluster core is due to the shock heating of preheated
gas. Furthermore, we have confirmed that the observed fundamental plane
and band of clusters are reproduced by the model even when the effects
of preheating are taken into account. Thus, major conclusions about the
cluster formation and cosmology obtained in our previous papers are not
changed.

\acknowledgments

We thank for A. C. Edge, C. S. Frenk, and T. Kodama for useful
discussions. This work was supported in part by the JSPS Research
Fellowship for Young Scientists.

\clearpage

\newpage

\newpage

\begin{figure}
\epsscale{1.00}
\plottwo{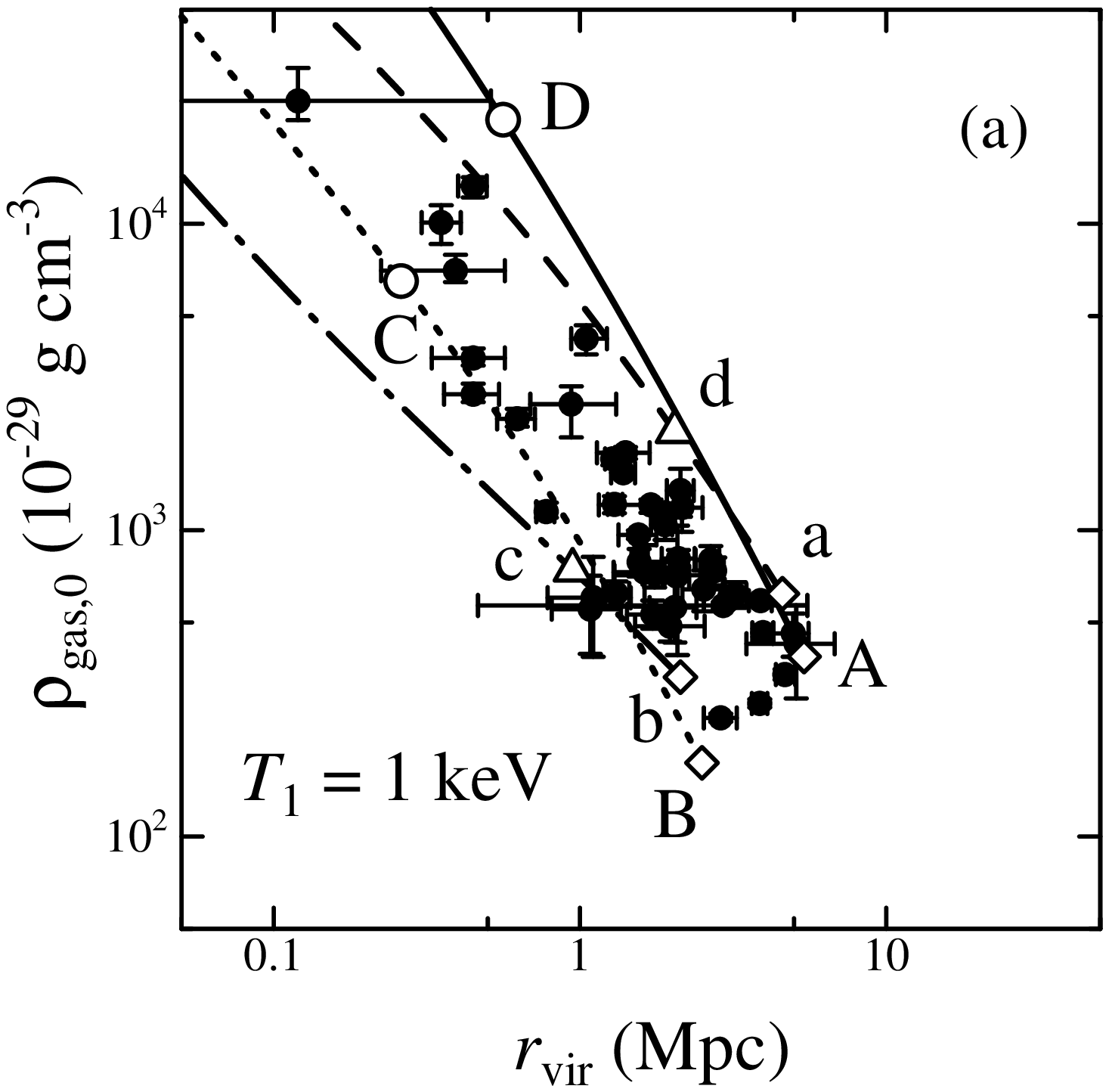}{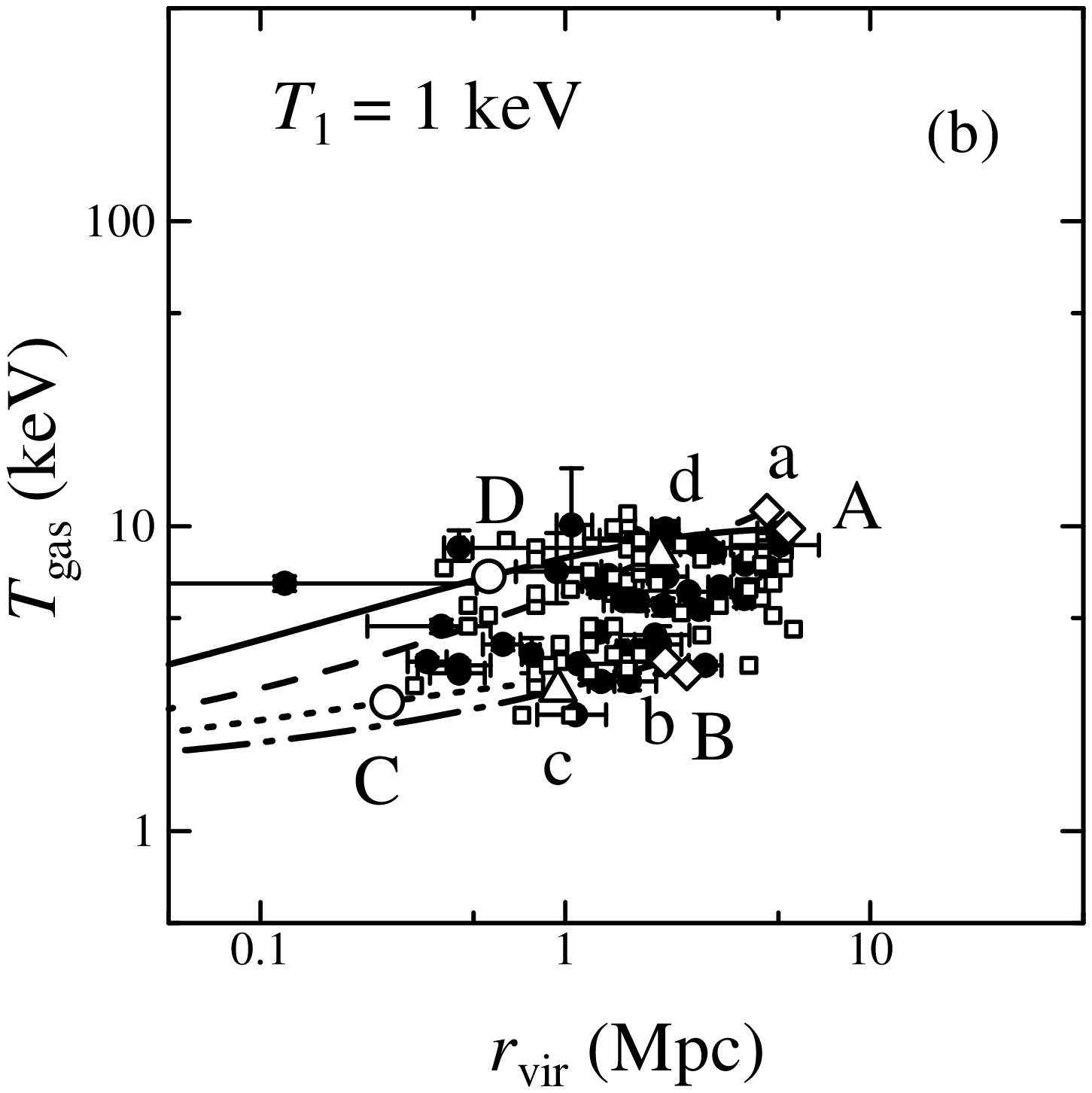}
\end{figure}

\begin{figure}
\figurenum{5}
\epsscale{0.45}
\plotone{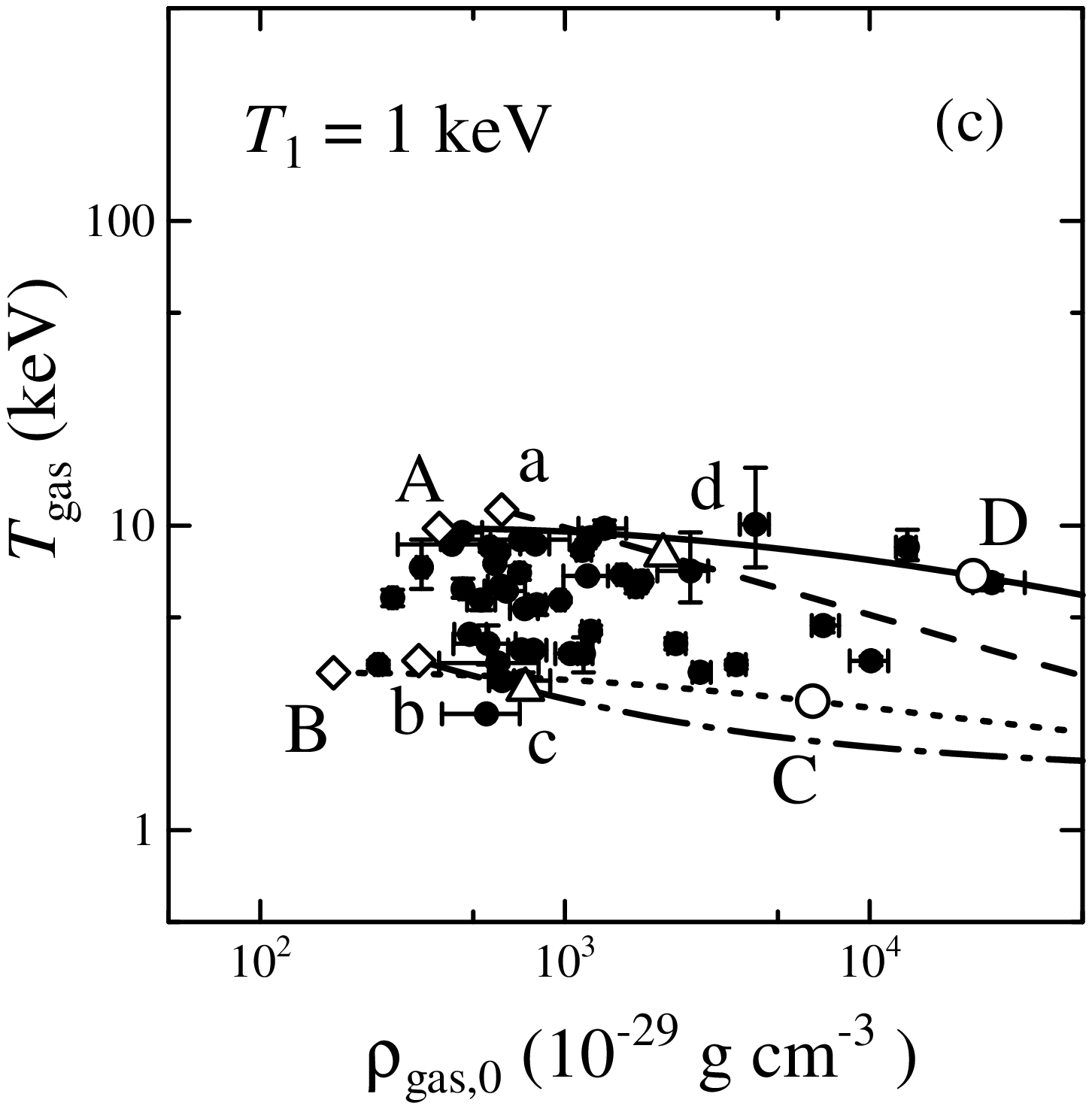}
\caption{Theoretical predictions of (a) radius--density
relation (b) radius--temperature relation (c) density--temperature
relation when $\gamma=1$. Solid line: $\Omega_0=0.2$ and $M_{\rm vir,
0}=5\times 10^{15}\rm\; M_{\sun}$. Dotted line: $\Omega_0=0.2$ and
$M_{\rm vir, 0}=5\times 10^{14}\rm\; M_{\sun}$. Dashed line:
$\Omega_0=1.0$ and $M_{\rm vir, 0}=5\times 10^{15}\rm\;
M_{\sun}$. Dash-dotted line: $\Omega_0=1.0$ and $M_{\rm vir, 0}=5\times
10^{14}\rm\; M_{\sun}$. The open diamonds, triangles, and circles
correspond to the collapse redshifts of $z_{\rm coll}=0$, $z_{\rm
coll}=0.5$, and $z_{\rm coll}=4$, respectively. We assume that $r_{\rm
vir}=8r_{\rm c}$. The observational data obtained by \citet{moh99}
(filled circles) and \citet{per98} (open squares, only in Figure
\ref{FB}b) are overlaid. \label{FB}}
\end{figure}

\newpage

\end{document}